\RequirePackage{fix-cm}
\documentclass[smallextended]{svjour3}       
\smartqed  
\usepackage{graphicx}
\usepackage{units}
\usepackage{subfigure}
\usepackage[usenames,dvipsnames]{color}
\usepackage{amsmath}

\newboolean{show_examples}
\setboolean{show_examples}{true}	
\newcommand{\myexample}[1]{\ifthenelse{\boolean{show_examples}}{#1}{}}

\graphicspath{{figures/}}

\begin{document}

\title{Quantitative Characterization of Components of Computer Assisted Interventions
}


\titlerunning{Quantitative Characterization of Components of CAI}

\author{Asl\i{}~Okur \and Ralf~Stauder \and Hubertus Feussner \and Nassir~Navab}
\authorrunning{A. Okur, R. Stauder, H. Feussner and N. Navab}   
\institute{ * Joint First Authors \\ \\
			A. Okur*, R. Stauder* and N. Navab \at
			Computer Aided Medical Procedures, Technische Universit\"at M\"unchen, Germany \\
			\email{asli.okur@tum.de, ralf.stauder@tum.de} 
			\and
			R. Stauder* and H. Feussner\at
			MITI, Klinikum Rechts der Isar, Technische Universit\"at M\"unchen, Germany 
			\and
			N. Navab \at
			Computer Aided Medical Procedures, John Hopkins University, USA 
}

\date{Received: date / Accepted: date}

\maketitle

\begin{abstract}
	
\emph{Purpose.}
We propose a mathematical framework for quantitative analysis weighting the impact of heterogeneous components of a surgery.
While multi-level appoaches, surgical process modeling and other workflow analysis methods exist,
this is to our knowledge the first quantitative approach.
\emph{Methods.}
Inspired by the group decision making problem from the field of operational research, 
we define event impact factors, which combine independent and very diverse low-level functions. 
This allows us to rate surgical events by their importance.
\emph{Results.}
We conducted surveys with 4 surgeons to determine the importance of roles, phases and their combinations within a laparoscopic cholecystectomy.
Applying this data on a recorded surgery, we showed that it is possible to define a quantitative measure for deciding on acception or rejection of calls to different roles and at different phases of surgery. 
\emph{Conclusions.} 
This methodology allows us to use components such as expertise and role of the surgical staff
and other aspects of a given surgery in order to quantitatively analyze and evaluate events, actions, user interfaces or procedures.
\end{abstract}

\section{Introduction} 
\label{sec:introduction}

Inside an operating room (OR) a large number of interactions happen on a regular basis. 
With the introduction of more advanced technical equipment such as imaging or navigation systems, these interactions are becoming even more complex.
Every action inside the OR is majorly influenced by different sources, unique to each particular intervention,
such as the team constellation and individual characteristics of the staff members.
These influences can be very diverse, and quantifying them as a single measurement is not trivial.
This paper aims at introducing mathematical methods for impact calculation for events happening inside the OR,
and their use in analysis of surgical processes and related computer assisted solutions.

\begin{figure}[htb]
	\center
	\includegraphics[width=0.70\linewidth]{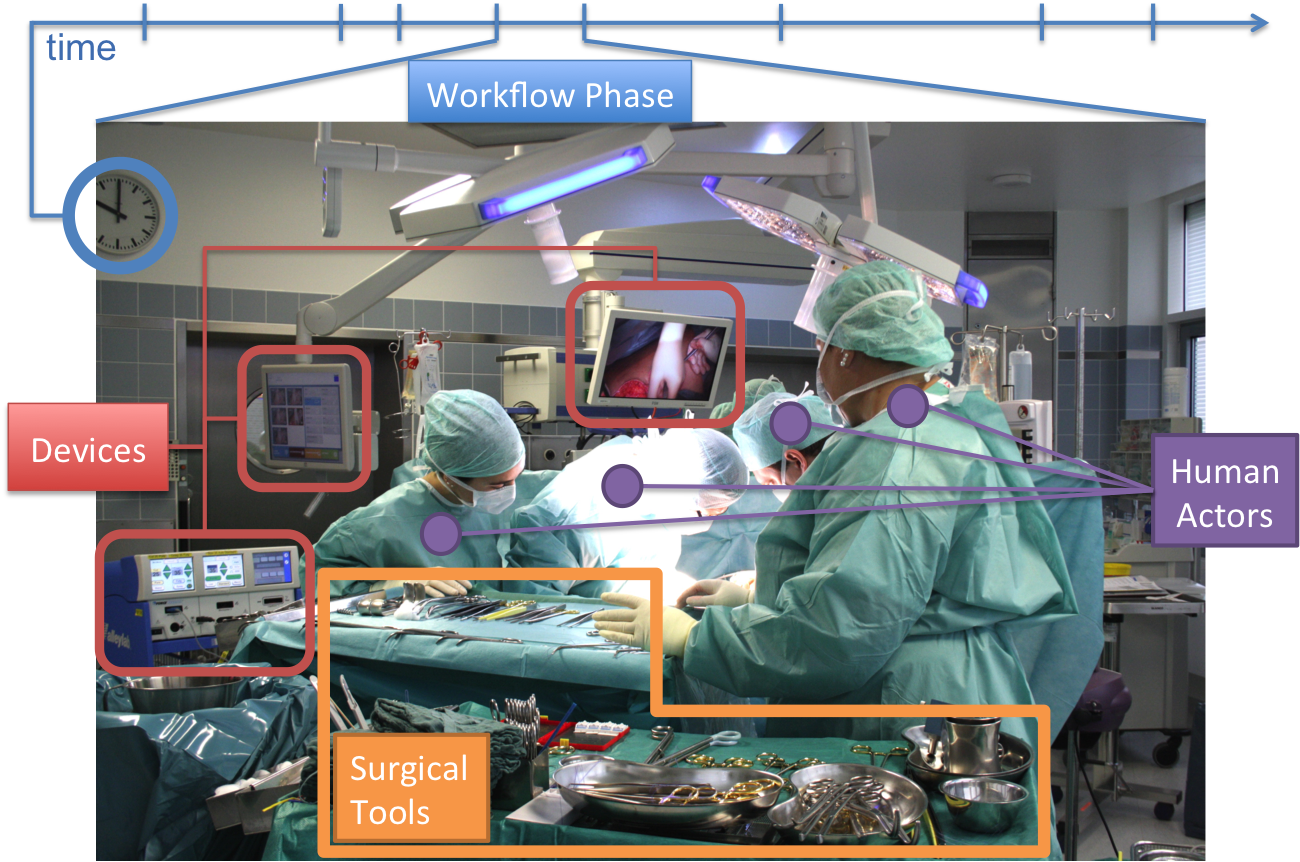}
	\caption{
		Actions inside the OR are influenced by different sources such as surgical workflow, human roles, devices and surgical tools. Image courtesy of Armin Schneider, MITI, TUM.
		}
	\label{fig:ORImage}
\end{figure}

Reaching decisions with multiple independent parameters has been under study for several years in operational research.
Diverse mathematical models have been developed to help groups decide and act in complex situations.
These methods have not yet been widely utilized in the medical domain, despite the importance of different factors when analyzing surgical processes.

A first step towards dividing a surgery into statistically reproducible phases and their characteristics are design and recovery of surgical process models \cite{Neumuth2011,Bouarfa2012a,Padoy2010}.
This analysis and its use in monitoring of the surgical workflow are important prerequisites to design and implement any advanced intraoperative system.
Jannin et al. \cite{jannin2008} introduced an assessment system for image guided interventions based on six levels.
Bigdelou et al. \cite{bigdelou2011ipcai} proposed a method for usability analysis of computer assisted surgery solutions
by defining items within different views in an OR specific domain model.
However, they do not provide a mathematical method allowing quantified analysis and in particular measuring the impact of different elements of such models.

In this work we bring a new methodology into the domain of computer assisted interventions
in order to determine the impact of surgical events based on different components of the OR.
The aim of the rating should be chosen according to an intended study by adjusting the component characteristic functions (CCF) introduced in \ref{sub:ccf}.
This enables many possible applications improving any aspect of a surgery such as patient outcome, surgery duration or healthcare costs. 
Here we chose a more immediate approach to minimize human error,
by evaluating whether a possibly disturbing incoming phone call should be rejected during any point of an intervention.


\section{Impact Calculation} 
\label{sec:impact_calculation}

The OR domain model proposed by Bigdelou et al. \cite{bigdelou2011ipcai} consists of three distinct \emph{views}:
\emph{surgical workflow}, \emph{target device} and \emph{human roles}.
\emph{Mappings} are defined as the correlation between two \emph{elements} of different \emph{views}.
These are represented in \emph{mapping tables}, which are useful for analyzing the complex connections between \emph{views}. 

Although this model provides a better analysis of the complex OR domain, 
it lacks a mathematical method allowing quantified analysis.
As some surgical workflow phases might be more critical than others or some human roles are much more important than the remaining ones,
an intelligent way to distinguish the influence of some \emph{view elements} on the surgery is required.
In this section we will introduce the methodology for impact determination for individual \emph{surgical events}
and clarify it with application examples. 

Please note that the methodology we use for impact calculation is inspired by the \emph{Group Decision Making Problem (GDM)},
which has been studied in the field of \emph{Operational Research} for more than 20 years now.
In GDM the opinions of different human experts are intelligently combined to a single, collective opinion to determine the best of several alternatives,
e.g. choosing as a group which university to study at, based on different, independent parameters such as costs, reputation, number of professors, social life etc.
The approach most suitable to our model was described in \cite{herrera_multiperson_2001}, while one of the first works in this field was done by Saaty \cite{Saaty1990}.
A very important factor in GDM are the employed similarity measures and combining operators, which are compared and described in \cite{Chiclana2012}. 
Usual GDM problems can be found in consensus systems and virtual communities \cite{Alonso2010}.
There have been some applications to the medical domain, but only in a very generic and limited fashion \cite{Perez2010}.

\subsection{Surgical Events} 
\label{sub:events}

An \emph{event} $e = (c_1, ..., c_l), l >= 2$ is a tuple of different \emph{components} 
$c$ describing the event, as seen from the different \emph{views}.
The whole surgery can be seen as the finite set $E = \{e_1, ..., e_n\}, n >= 2$ of all \emph{events} that happen within the surgery.
\myexample{
For example an event might be
the \emph{circulator} providing a new pair of \emph{surgical gloves} during the \emph{drainage} phase of a surgery,
or the \emph{main surgeon} using the \emph{laparoscopic scissors} during the \emph{clipping and cutting} phase.
}

The specific definition of an \emph{event} stongly depend on the intended study. 
An exhaustive option would be to define an event as any distinct action that is done by a person in the OR using any tool or device.
Alternatively only feedback given directly by surgical staff members can be considered \emph{events}.

Lalys et al. \cite{Lalys2014} define several aspects to differentiate surgical workflow analysis methods.
Among other aspects they mention the level of granularity,
for which they describe a full spectrum with low-level information such as video frames on the lowest end and the whole procedure at the top-most end.
With regard to our method, anything recorded with a granularity of Activities or Steps (to a lesser degree also Motions) is likely suitable to be seen as \emph{surgical event},
while the defined Phases (and possibly the whole Procedure) can be used as a \emph{component}. 

In \cite{Haro2012} Haro et al. detect surgical gestures through identifying \lq\lq{}surgemes\rq\rq{} (which are equivalent to Activities in \cite{Lalys2014}) in video data.
These detected Activities can easily be used as \emph{events} in our work.
Also when using a collaborative human-robot environment as described by Padoy et al. \cite{Padoy2011},
both the detected manual subtasks (which can be compared either to Activities or Motions in~\cite{Lalys2014}) and the subtasks performed automatically by the robot
are suitable to be considered \emph{events}.

Kranzfelder et al. \cite{Kranzfelder2013} attached RFID chips to surgical instruments to recognize their usage and identify a surgical step from that information.
Depending on the intended insight there are several ways to apply our method to the data provided in their work:
One option is to use the detected usage of an instrument as \emph{events}, while the predefined steps can be seen as a \emph{component} (in parallel to the \emph{instruments component}). 
This would yield a very detailled and exhaustive list of \emph{events} per intervention with a rather flat distribution of relative impact per event.
Another way could be the opposite and interpret the mentioned steps as \emph{events},
while the instruments used within the steps contribute \emph{components} as an \emph{instrument view}.
This will give less \emph{events}, but their respective impact should be more distinctive.

Usability feedback like comments and complaints by the OR staff about a single medical device was recorded for the experiments in the work by Bigdelou et al. \cite{bigdelou2011ipcai}.
In their work the comments are filtered and sorted according to the connected \emph{views} (like workflow phase or human role),
while these recorded feedback elements would represent \emph{events} in our method.
Each of these \emph{views} can be used as a \emph{component} for event impact calculation (in this case feedback impact calculation).
Also the work of Neumuth et al. \cite{Neumuth2012} records surgeries in a very detailed way.
The surgical work steps (or Activities) mentioned in their work correspond to \emph{events} in our method while the various perspectives can be translated to \emph{components}.
Therefore the generated patient-individual surgical process models (iSPMs) are directly suitable for use in our method.


\subsection{Component Characteristic Functions (CCF)} 
\label{sub:ccf}

The \emph{components} of each \emph{event} correspond to the different \emph{views} that are employed for analyzing the surgery.
As each individual \emph{view element} is usually defined as a free, text-based label,
we need functions for each \emph{component} to turn these labels into numeric \emph{characteristics}.
Also every \emph{component} can possess multiple \emph{characteristics} depending on the interpretation.

\myexample{
We will provide examples for better understanding here and throughout the article.
For the \emph{component} \emph{human role} of an \emph{event} happening inside the OR, we define two \emph{characteristics}:
experience of the person (e.g. 15 years) and
a rating of the influence of that person's role on the medical outcome of the surgery (based on a survey among medical experts).
}

For some \emph{components} direct measurements can be obtained easily,
while for other \emph{components} or \emph{characteristics} 
this is more subjective and can only be determined via expert surveys.
Therefore we introduce three different types of \emph{Component Characteristic Functions (CCF)}:
\emph{CCF-Ordering (CCFO)}, \emph{CCF-Rating (CCFR)} and \emph{CCF-Pairwise Comparison (CCFP)}.\footnote{
More detailed explanations, further such structures and their application in group decision making problems can be found in \cite{herrera_multiperson_2001,Alonso2010}.}

\subsubsection{Component Characteristic Function ``Ordering'' (CCFO)} 
\label{ssub:ccf-o}

This is the least discriminative among the mentioned three \emph{CCF} types,
but usually the easiest to obtain when it is possible to define a total ordering over all \emph{characteristics}.
A \emph{CCFO} can be defined as $\text{\emph{O}} = \{o(1), ... , o(n)\}$,
with $o(\cdot)$ being a permutation over the index set $\{1, ... , n\}$ of all possible \emph{events}.
An ordering value of 1 designates the best ranked \emph{event}, based on the \emph{characteristics} for the considered \emph{component}. Based on the motivation of the study the meaning of ``best'' can be considered e.g. as ``fastest'', ``cheapest'' or ``safest''.

\myexample{
For example, a group of surgical experts might rank the phase \emph{preparation of Calot's Triangle} of a laparoscopic cholecystectomy as the most important \emph{surgical workflow} phase (giving it rank 1),
while the \emph{closure} phase might be considered less important (resulting in a low rank, e.g. 7).
}

An ordering is the simplest and fastest \emph{CCF} to be created manually.
This can easily be done through surveys with medical experts, in which they rank the \emph{components} in question based on their subjective opinion.
As this requires only comparisons of \lq\lq{}more\rq\rq{} and \lq\lq{}less important\rq\rq{},
even a relatively large number of elements can quickly be sorted and turned into a \emph{CCFO}.


\subsubsection{Component Characteristic Function ``Rating'' (CCFR)} 
\label{par:ccf-r}

A \emph{CCFR} calculates a specific utility value $u$ as \emph{characteristic} for a given component of each \emph{event}.
The function is defined as $\text{\emph{U}} = \{u(e_1), ... , u(e_n)\}$.
Higher utility values indicate a higher importance, but they do not have to be normalized\footnote{
Normalizing the utility values without loss of generality to $]0, 1]$ helps proving the properties of the transformation functions in \cite{herrera_multiperson_2001}.}.
If measurements of any kind are available (e.g. costs, power consumption) these can be used directly as utility value for a \emph{characteristic}.
Compared to \emph{CCFOs}, an advantage of \emph{CCFRs} is their ability to provide more information about the relative difference between pairs of \emph{events}.
While a \emph{CCFO} does not define a distance between neighboring \emph{components},
the same \emph{components} can have very different values in a \emph{CCFR} and therefore very diverse distances between them.
Conversely this also allows the definition of ties among \emph{components} by assinging them the same \emph{characteristic} value.
Each utility value $u$ for a \emph{CCFR} can be calculated independently of other values or \emph{events}.

\myexample{
As already stated in the examples above, the \emph{human role} could be valued by the experience in years of the involved persons
(e.g. 1 for a nurse in their first year or 20 for a resident surgeon after two decades of experience).
For the phases of a surgery a survey could ask experts of the OR domain such as surgeons to rate individual phases independently on a scale from 1-10,
instead of ranking them relative to each other,
e.g. giving the \emph{preparation} phase a value of 9 indicating high importance, and the \emph{closing} phase a value of 3.
It is also possible to rate two phases equally, e.g. giving both the \emph{gallbladder detachment} and the \emph{gallbladder retrieval} phase a value of 6,
and therefore giving them exactly the same importance, which is not possible with a \emph{CCFO}.
}


\subsubsection{Component Characteristic Function ``Pairwise Comparison'' (CCFP)} 
\label{subs:ccf-p}

\emph{CCFP} offers the highest level of detail,
as it also allows intransitive relations and relative importance loops between different components.
A \emph{CCFP} can be obtained by directly comparing all $\frac{(n-1)n}{2}$ pairwise combinations of the $n$ \emph{events}.
A $\text{\emph{CCFP}}$ is presented in a $n \times n$ unitriangular matrix $\mathbf{P}$.

Traditionally a very fine survey scale is used for the comparisons.
Saaty \cite{Saaty2008} gives up to 9 degrees from \emph{1 - equal importance} to \emph{9 - extreme importance} of the first \emph{event} over the second,
with the reciprocals ($\nicefrac{1}{1}$ through $\nicefrac{1}{9}$) depicting preference of the second \emph{event} over the first.
Other scales have been proposed with 5 or 7 steps, based on the psychological theory that human beings can usually handle $7\pm2$ information facts at once.
One could also think of a 11 step scale to represent a range from $[-5, 5]$.
In any case these values will be normalized to a common range in the following steps.
The scale of 9 steps used in this paper is based on a smaller scale of 5 steps as defined in \cite{Saaty1990},
referring to the informal judgements \lq\lq{}equal\rq\rq{}, \lq\lq{}moderately more\rq\rq{}, \lq\lq{}strongly more\rq\rq{},
\lq\lq{}very strongly more\rq\rq{} and \lq\lq{}extermely more\rq\rq{} (1, 3, 5, 7, 9), with added compromise values in between (2, 4, 6, 8).



\subsection{Component Characteristics Matrix (CCM)} 
\label{sub:ccm}

After obtaining different \emph{CCFs} from various sources, we need to unify the collected information.
In order to retain the highest possible level of detail,
we will transform all individual \emph{CCFs} to \emph{component characteristic matrices (CCM)}.
First we will define the structure of a \emph{CCM}, which enables the further calculations in our model,
and later the transformation functions needed for each type of \emph{CCF}.

A \emph{CCM} is a matrix $\mathbf{M} \subset n \times n$,
in which every element $m_{ij}$ represents the relative importance of event $e_i$ over event $e_j$.
$\mathbf{M}$ is a multiplicative reciprocal matrix,
i.e. for every element $m_{ij}$ the condition holds that $m_{ij} \cdot m_{ji} = 1$.
The values of $\mathbf{M}$ are bound to $[\nicefrac{1}{9}, 9]$, comparable to typical values for a \emph{CCFP}.
As above, a value of 1 represents indifference between the two compared events,
a value of 9 indicates a high relative importance of event $e_i$ over $e_j$
and a value of $\nicefrac{1}{9}$ indicates the opposite.

\subsubsection{From CCFO to CCM} 
\label{ssub:ccf_o2ccm}

We are looking for a function $f$ that transforms a $\text{\emph{CCFO}}$ into a $\text{\emph{CCM}}$.
Every value $m_{ij}$ of the matrix $\mathbf{M}$ is linked to exactly two events $e_i$ and $e_j$,
so a suitable transformation function must use the two corresponding ordering values as parameters $o(i)$ and $o(j)$.
A lower value of $o(\cdot)$ denotes more importance over a higher value.
The difference between the places of the ordering can express the degree of influence,
so a function based on the difference of ordering values is most appropriate,
$
	m_{ij} = f(o(i), o(j)) = g(o(j) - o(i)).
	\label{eq:ccf-o2ccm_def}
$

To ease further calculations and based on the proof of \cite{herrera_multiperson_2001}, we replace the ordering values by inverted, normalized substitute values
$
	s_i = \frac{n - o(i)}{n-1},
	\label{eq:ccf-o2ccm_subst}
$
which also replaces the difference of ordering values by the difference $s_i - s_j$.
The general solution for functions based on the difference of parameters which fulfill our additional requirements are exponential functions \cite{herrera_multiperson_2001},
so our final transfer function, using the substitution values from above, is
\begin{equation}
	m_{ij} = f(o(i), o(j)) = 9^{s_i - s_j}.
	\label{eq:ccf-o2ccm_final}
\end{equation}


\subsubsection{From CCFR to CCM} 
\label{ssub:ccf_rating_to_ccm}

We also need a function $h$ to transform a rating $\text{\emph{CCFR}}$ to a $\text{\emph{CCM}}$.
Again the values $m_{ij}$ of the matrix have to be computed from two utility values of the compared events
$
	m_{ij} = h(u(e_i), u(e_j)).
	\label{eq:ccf-r2ccm_def}
$
A bigger difference between the values $u(e_i)$ and $u(e_j)$ indicates a stronger influence of $e_i$ over $e_j$.
$h$ should be a continuous function, increasing in the first parameter and decreasing in the second.
In order to follow the multiplicative reciprocity constraint, the function must also fulfill
$
	h(x, y) \cdot h(y, x) = 1.
	\label{eq:ccf-r2ccm_prop}
$
The ratio between the utility values can be expressed within a family of functions that satisfy these conditions:
\begin{equation}
	m_{ij} = h(u(e_i), u(e_j)) = \left( \frac{u(e_i)}{u(e_j)} \right)^z.	
	\label{eq:ccf-r2ccm_final}
\end{equation}
Choosing the trivial value of $z = 1$ is a valid option and will be used throughout the remainder of this paper.
Higher values of $z$ can be chosen to enhance the influence of higher ratios and amplify large differences between utility values.
Alternatively one can use smaller values for $z$ to dampen this effect and achieve more adjacent ratios even for large utility value differences.
The full proof and description of this function family is given in \cite{herrera_multiperson_2001}.

Independent of the actual range of the utility values,
this ratio can easily exceed the expected value range of $[\nicefrac{1}{9}, 9]$ for the \emph{CCM},
so we need to normalize the matrix afterwards.
Let the transformed rating be in the value range of $[\nicefrac{1}{m}, m]$.
The normalization must maintain the reciprocity of the matrix, so a suitable function is
$norm(x) = x^{1/log_9(m)}$.
This function is applied to every value of \emph{M}.
Further description of the properties of this function and the proof are available in \cite{Herrera-Viedma2004}.


\subsubsection{From CCFP to CCM} 
\label{subs:ccf-p2ccm}

A \emph{CCFP} is usually already presented in an $n \times n$ unitriangular matrix $\mathbf{P}$ with a suitable range of $[\nicefrac{1}{9}, 9]$.
Therefore a transformation to a \emph{CCM} can be achieved by simply filling the rest of the matrix by exploiting the multiplicative recoprocity:
\begin{equation}
	m_{ij} = \frac{1}{m_{ji}}.
	\label{eq:ccf-p2ccm_final}
\end{equation}



\subsection{Collective Component Characteristics Matrix (CCCM)} 
\label{sub:combined_scoring_matrix}

After transforming all available CCFs to CCMs $\{\mathbf{M}^1,..., \mathbf{M}^r\}$,
we combine all to a \emph{Collective Component Characteristics Matrix (CCCM)} $\mathbf{M}^c$.
Each value of $\mathbf{M}^c$ is the geometric mean of the corresponding value of all individual matrices,
\begin{equation}
	m_{ij}^c = \phi^G(m_{ij}^1,..., m_{ij}^r) = \prod_{k=1}^r(m_{ij}^k)^{1/r}.
	\label{eq:col_aggregation}
\end{equation}
The geometric mean could be replaced by any \emph{ordered weighted geometric (OWG)} operator,
including fuzzy majority quantifiers as in \cite{herrera_multiperson_2001,Herrera-Viedma2004} 
or variations of the OWG or ordered weighted averaging (OWA) operators as given in \cite{Merigo2010}.


\subsection{Event Impact Factor (EIF)} 
\label{sub:weighting_vector}

Using the collective component characteristics matrix (CCCM) $\mathbf{M}^c$, we can calculate a scalar value for each \emph{event} that combines the influence of all \emph{characteristics} as following:
\begin{equation}
	I_i = \frac{1}{2} \cdot (1 + log_9\phi^G(m_{ij}^c, j=1,...,n))
	\label{eq:mqgdd}
\end{equation}
This is one of the two measures that is suggested in \cite{herrera_multiperson_2001} for ranking all the alternatives and for selecting the $s$ best ones,
which we again reduce to the geometric mean as special case of the OWG.
After normalization among all the \emph{events}, we use this vector directly as measurement of the relative \emph{impact} of all \emph{events} in our OR Domain Model.
Therefore we define the individual values as \emph{event impact factors (EIF)}. 



\section{Application to the OR Domain} 
\label{sec:application_to_the_or_domain}


To demonstrate the applicability in the OR of the methods described above,
we apply the methodology to a laparoscopic cholecystectomy.
The human roles are modeled using five actors: \emph{main surgeon (M.S.), assistant surgeon (A.S.), nurse (N.), circulator (C.)} and \emph{anesthetist (A.)}.
For the sake of simplicity, the surgical workflow is modeled as a linear flow chart consisting of seven workflow phases:
\emph{Trocar placement (Troc.), preparation of Calot's Triangle (Prep.), clipping and cutting of cystic duct and artery (Clip.),
detachment of the gallbladder from the liver bed (Det.), retrieval of the gallbladder (Retr.), check for bleedings and hemostasis (Hemo.)} and \emph{drainage and closing (Clos.)}.

The CCFs used in this study are explained below. We do not claim that these CCFs are exhaustive or ideal choices, although the achieved results shown in Section \ref{sec:results} demonstrate even few and simple CCFs may be sufficient.

\subsection{Meta-Components} 
\label{sub:meta_components}
\begin{figure}[tb]
	\center
	\includegraphics[width=0.90\linewidth]{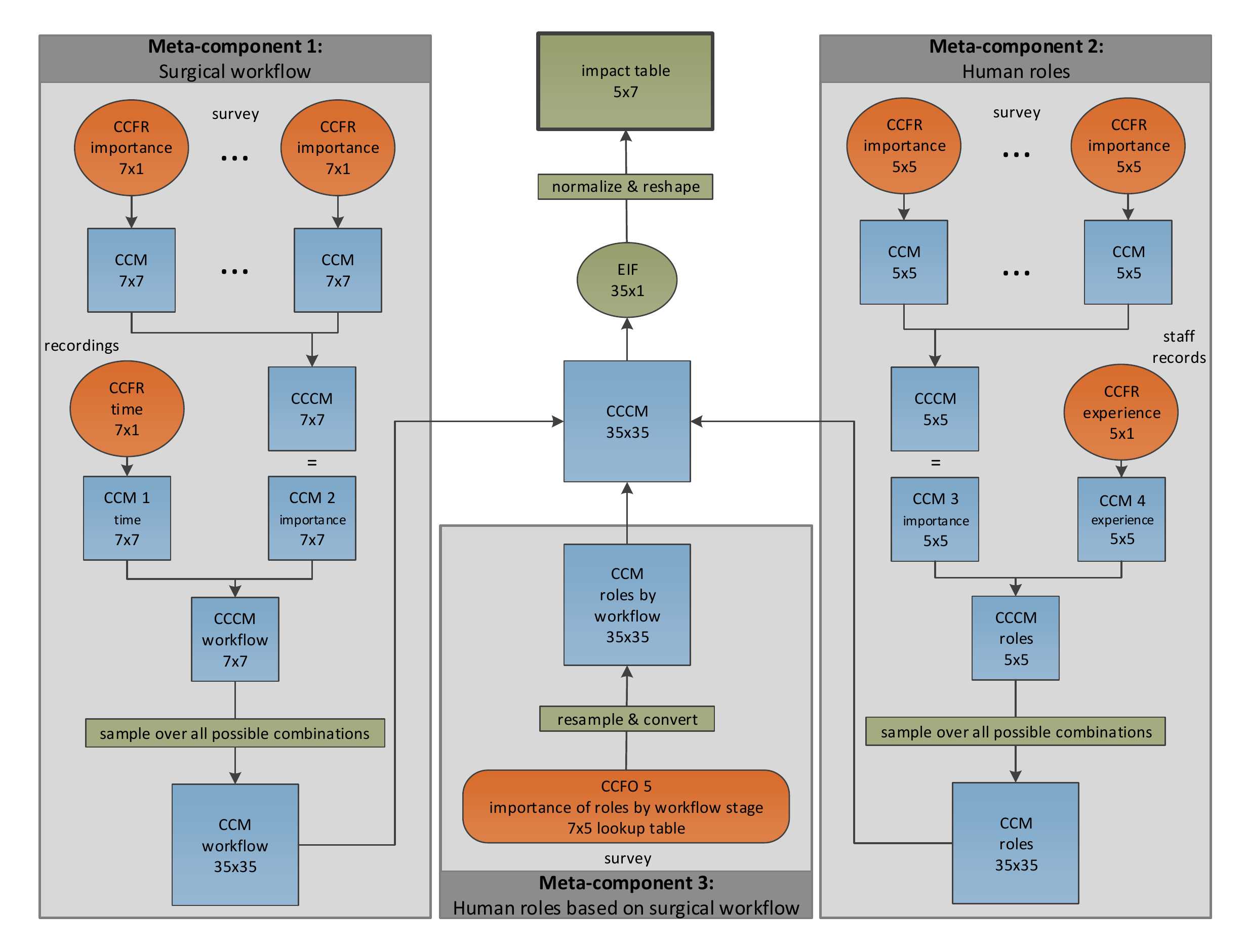}
	\caption{
		Complete data flow of our example. 
		}
	\label{fig:calculation_flow}
\end{figure}

We apply this methodology in different layers of our modeling pipeline (Fig.~\ref{fig:calculation_flow}).
Therefore we differentiate between \emph{components} and \emph{meta-components},
the latter being hierarchically above and derived from one or several \emph{components}.
We define three \emph{meta-components} in our model.

\subsubsection{Surgical Workflow} 
\label{subs:surgical_workflow}
We start with a simple \emph{meta-component}, which takes into account only one \emph{component}: Surgical Workflow. 
We define two different \emph{characteristics} for this meta-component.
The first \emph{characteristic} is \emph{time}. 
For each of our 7 defined workflow phases we calculated the average duration per phase over 7 pre-recorded and manually labeled surgeries.
The characteristic $CCF_1^{w}$ can be represented as a CCFR defined by the duration of each workflow phase in seconds.
The last phase \emph{drainage and closing} was in average the shortest phase with 172s, while the fourth phase \emph{dissection of the gallbladder} took the longest with an average of 562s.
Based on this data we will use the utility function $U^1 = \{179, 419, 390, 562, 390, 337, 172\}$.
This CCF is then transformed using eq.~\ref{eq:ccf-r2ccm_final} to a $7 \times 7$ $CCM_1^{w}$.

The second characteristic $CCF_2^{w}$ for this meta-component is itself a $7 \times 7$ CCCM based on questionnaires filled out by four medical experts.
We asked four experts to give each phase a score from 1-10 (implementing a CCFR), rating each phase by their respective influence on the total medical outcome.
We then converted these CCFRs from each survey to CCMs using eq.~\ref{eq:ccf-r2ccm_final}, combined them to a $CCCM^{w}_{survey}$ through eq.~\ref{eq:col_aggregation},
and used the resulting matrix directly as $CCM_2^{w}$ for all further calculations.

Finally the collective $7 \times 7$ component characteristics matrix $CCCM^{w}$ is calculated
by applying eq.~\ref{eq:col_aggregation} on all $CCM^w$.


\subsubsection{Human Role} 
\label{subs:human_role}
This meta-component function only rates the human roles and is not influenced by the surgical workflow.
Here we define again two characteristics, which will be used later to calculate the CCCM for this meta-component. 
The first one $CCF_3^{r}$ is the \emph{importance of the role}, which can be determined as above by user surveys as a $5 \times 5$ CCCM.
We asked four medical experts to also rate the influence of a specific role (e.g. assistant surgeon or scrub nurse) on the medical outcome, on the same scale of 1-10.
The collected CCFRs were again converted and combined as above to directly yield $CCM_3^{r}$.

However, this characteristic does not take into account the influence of different individuals involved in the surgery, but only considers the role itself.
Therefore we introduced an additional characteristic $CCF_4^{r}$, which measures the person\rq{}s \emph{experience} in years.
This value can be used directly as a utility function, e.g. $U^4 = \{30, 1, 1, 5, 10\}$ in our case.

The $5 \times 5$ $CCCM^{r}$ is calculated by applying eq.~\ref{eq:col_aggregation} on all $CCM^r$.


\subsubsection{Human Roles Based on Surgical Workflow} 
\label{subs:human_roles_based_on_surgical_workflow}
Throughout the whole surgery, the importance of a specific role might be comparably low,
however in some specific phases the person can have the most relative importance (like the \emph{scrub nurse} during the \emph{trocar placement} phase).
This can be modeled easily using a simple CCF like ordering the roles for each workflow phase separately,
which were again obtained through expert surveys.
For example, for the $CCF_5^{rw}$ for the phase \emph{trocar placement} one ordering was $O^5_{Troc.} = \{1, 4, 2, 5, 3\}$,
while during the \emph{gallbladder retrieval} we had an ordering of $O^5_{Retr.} = \{1, 2, 3, 5, 4\}$.
By applying the equations \ref{eq:ccf-o2ccm_final} and \ref{eq:mqgdd} to each CCF, we get utility values for a rating function $CCFR_5^{rw}$,
which we can store in a $7 \times 5$ look-up table.



\subsection{Final Impact Calculation} 
\label{sub:final_impact_calculation}

Until now the meta-components each provided CCFs or CCMs in different sizes.
Therefore, we need to resample the information and convert them to CCMs of the same size,
so that we can apply aggregation and impact calculation as described in Section~\ref{sec:impact_calculation}.
For every combination of workflow phase and human role we create a single, virtual \emph{event},
for which we evaluate every meta-component accordingly to build the $35 \times 35$ collective CCM through eq.~\ref{eq:col_aggregation}.
The $35 \times 1$ value vector obtained by eq.~\ref{eq:mqgdd} is then normalized and determines the final event impact factors.
The event impact factors are then reordered to a $5 \times 7$ look-up table for easier visualization (Fig.~\ref{fig:results}).



\section{Results: Proof of Concept} 
\label{sec:results}

\begin{figure}[tb]
	\centering
	\subfigure[Impact table normalized over all entries] {
		\includegraphics[width=0.46\linewidth]{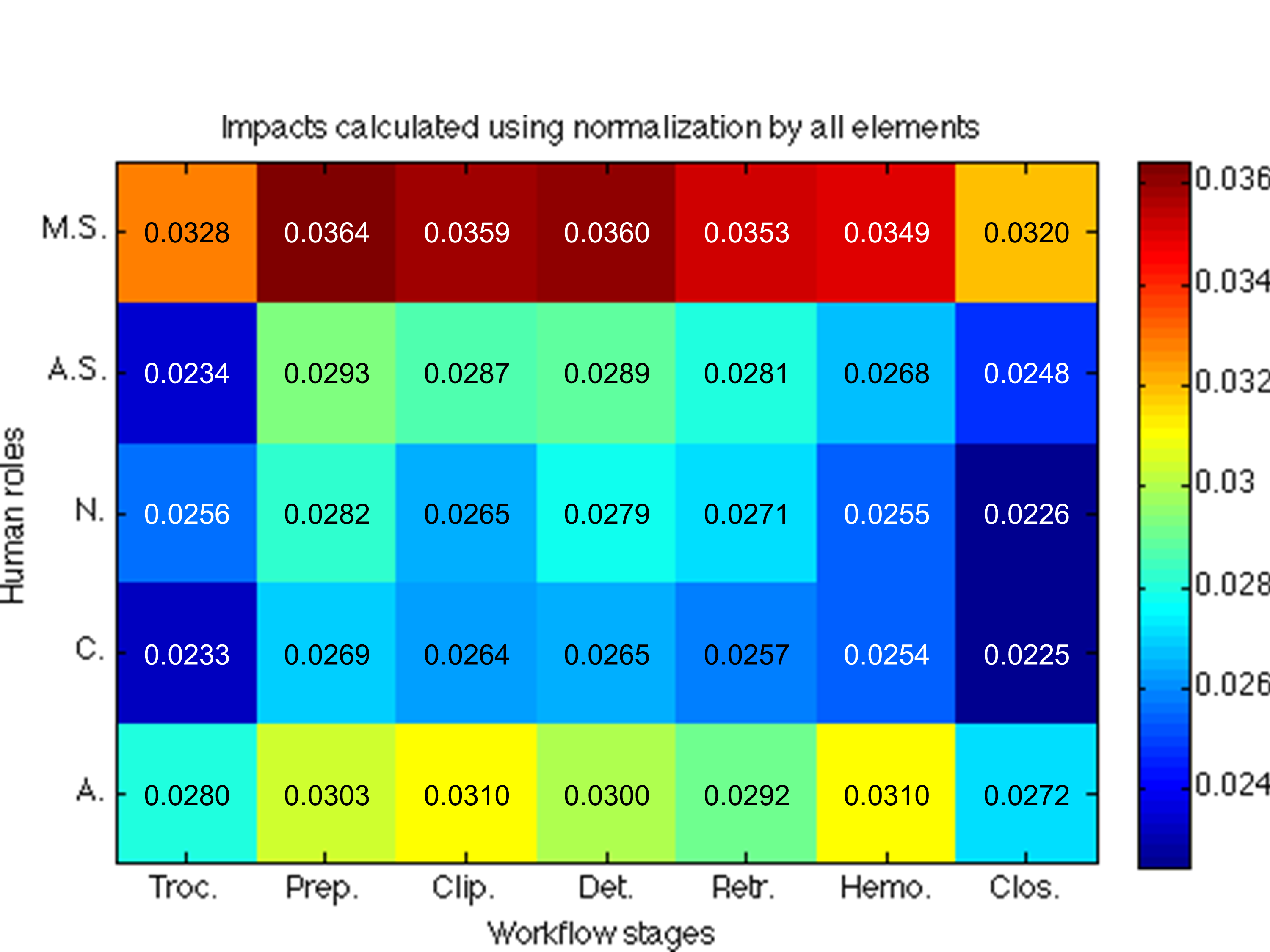}
		\label{fig:col_weight_norm}
	}
	\subfigure[Impact table with trainee and experienced surgeon switching roles]{
		\includegraphics[width=0.46\linewidth]{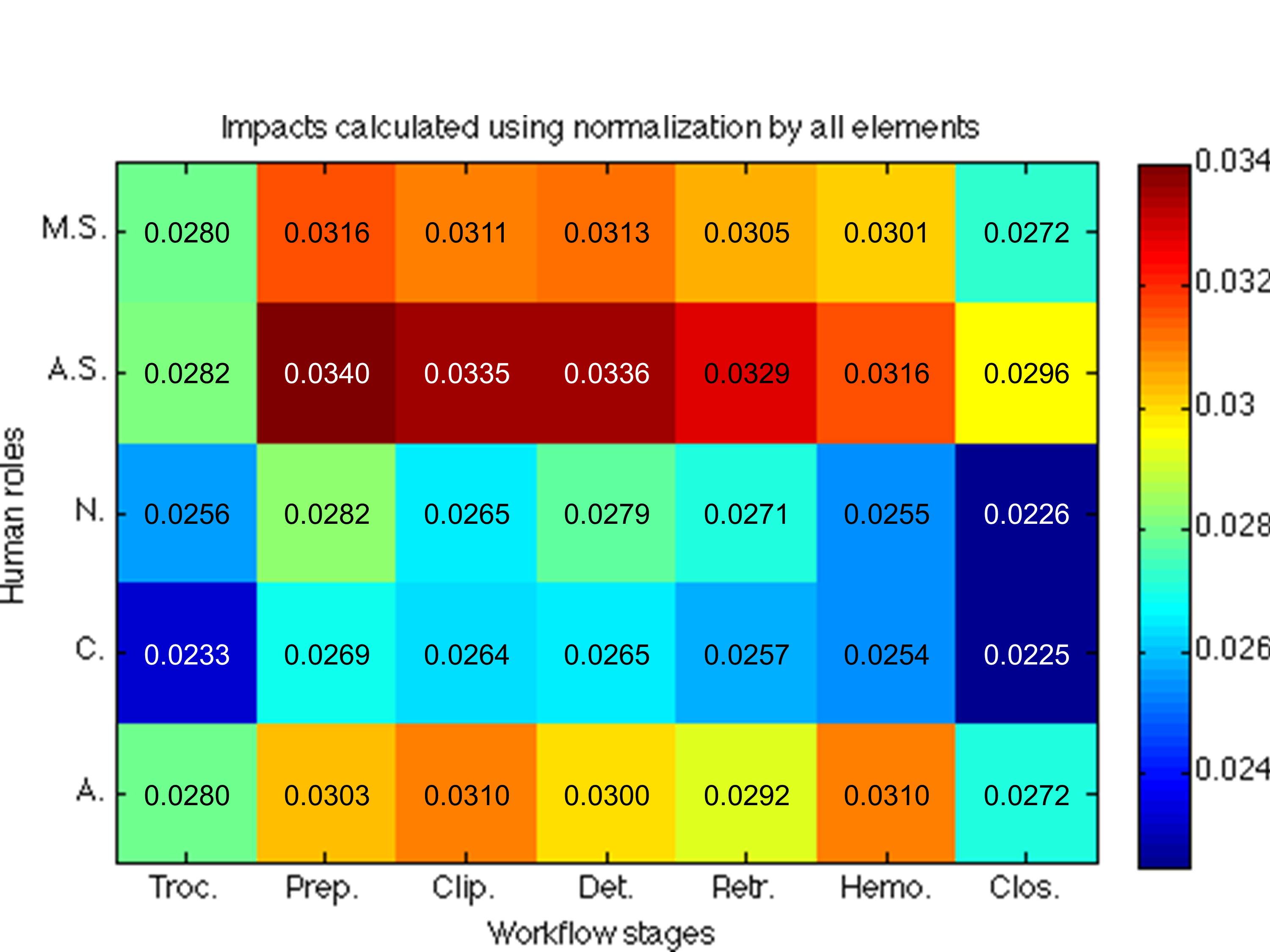}
		\label{fig:col_changed}
	}
	\caption{The final impact tables for the recorded surgery, by workflow and role.
	}
	\label{fig:results}
\end{figure}

After obtaining the final event impact factors (EIF), we have several opportunities to apply it to the OR.
A first application is for device manufacturers to collect usability feedback for a prototype during surgeries and rank the feedback items by their importance.
As these feedback items are connected to the workflow phase and the person who provided them,
ranking them can be done by simply looking up the corresponding value in our impact table (see Fig.~\ref{fig:col_weight_norm}). 
This way, instead of doing a time consuming and expensive manual review of all feedback items first,
developers can immediately focus on the most critical elements, identified by the highest impact factors associated with them.

Our proposed method is very flexible and inherently adjusts to changes in the environment.
Given two interventions of the same type, the chief physician (experience of 30 years) acts as the main surgeon in the first,
but promotes the less experienced trainee physician (in their first year) to this role in the second.
This only affects one of the two human role characteristic functions described in \ref{subs:human_role},
however the resulting impact vector and impact table change significantly (see Fig.~\ref{fig:col_changed}).
Now the highest impact is to some extent split between the \emph{role} main surgeon and the \emph{experience} of the person chief surgeon,
who took the role of assistant surgeon. 
This allows the comparison of a wide range of elements across various team constellations and even surgical schools. 

Phone calls during critical phases of a surgery can severely distract the surgical crew.
However they could be crucial or of high urgency for one of the players in OR.
It is important to quantify and classify them in regard to their importance, the phase of surgery and the person the call is destinated to.
An automatic call handling system could decide to accept or reject the call to the person based on the importance of the current phase, which can be quantified using event impact factors. 
If the combined impact factor of the phase and called person is higher than a defined threshold,
the current phase is too crucial to be interrupted and all phone calls to that person should be blocked.
The highest impact factor in our case was 0.0364 for the combination of \emph{main surgeon} and \emph{preparation of Calot's Triangle} phase (See Fig.~\ref{fig:col_weight_norm}).
Based on the discussions with our expert surgeon, we identified 98\% of the maximum event impact factor to be a suitable threshold 0.0357 for our examined surgery.
Therefore the system would automatically refuse calls for the main surgeon in phases \emph{Prep.}, \emph{Clip.}, and \emph{Det.} (corresponding to impact factors 0.0364, 0.0359 and 0.0360).
This correlated well with the strategy of call transfer proposed by the chief surgeon, who had no access to this data.

This phone call handling system can also be extended by including additional information about the call itself to the \emph{EIF} calculation through further CCFs,
to be able to better distinguish between regular and urgent calls.
Some possible \emph{components} could be the caller, so that calls from predefined VIPs have a higher importance,
or the number of times a single caller already called in a predefined timeframe, increasing the chance of urgent calls over time.


\section{Conclusion} 
\label{sec:conclusion}

In order to include several, very different aspects of a surgical model into one combined impact table,
we applied methods from the field of operational research to the medical domain.
To the best of our knowledge, this methodology was utilized here for the first time in our community.
We provided an example of application of such concept and suggested several promising areas where such methods can enable quantitative analysis. 
Defining suitable models and component characteristic functions, finding more application areas, and obtaining well-founded values based on these calculations are next steps ahead.
This will be the challenge we offer to the community for the coming years.



\bibliographystyle{spmpsci}      
\bibliography{bibliography}

\begin{thebibliography}{10}
\providecommand{\url}[1]{{#1}}
\providecommand{\urlprefix}{URL }
\expandafter\ifx\csname urlstyle\endcsname\relax
  \providecommand{\doi}[1]{DOI~\discretionary{}{}{}#1}\else
  \providecommand{\doi}{DOI~\discretionary{}{}{}\begingroup
  \urlstyle{rm}\Url}\fi

\bibitem{Alonso2010}
Alonso, S., Herrera-Viedma, E., Chiclana, F., Herrera, F.: {A web based
  consensus support system for group decision making problems and incomplete
  preferences}.
\newblock Information Sciences \textbf{180}(23), 4477--4495 (2010)

\bibitem{bigdelou2011ipcai}
Bigdelou, A., Sterner, T., Wiesner, S., Wendler, T., Matthes, F., Navab, N.:
  {OR Specific Domain Model for Usability Evaluations of Intra-operative
  Systems}.
\newblock In: R.~Taylor, G.Z. Yang (eds.) IPCAI 2011, \emph{LNCS}, vol. 6689,
  pp. 25--35. Springer (2011)

\bibitem{Bouarfa2012a}
Bouarfa, L., Dankelman, J.: {Workflow mining and outlier detection from
  clinical activity logs}.
\newblock J. Biomed. Inform. \textbf{45}(6), 1185--90 (2012)

\bibitem{Chiclana2012}
Chiclana, F., {Tapia Garc\'{\i}a}, J.M., del Moral, M.J., Herrera-Viedma, E.:
  {A statistical comparative study of different similarity measures of
  consensus in group decision making}.
\newblock Information Sciences pp. 1--19 (2012)

\bibitem{Haro2012}
Haro, B.B., Zappella, L., Vidal, R.: {Surgical Gesture Classification from
  Video Data}.
\newblock In: MICCAI, pp. 1--8 (2012)

\bibitem{herrera_multiperson_2001}
Herrera, F., Herrera-Viedma, E., Chiclana, F.: Multiperson decision-making
  based on multiplicative preference relations.
\newblock Eur. J. Oper. Res. \textbf{129}(2), 372--385 (2001)

\bibitem{Herrera-Viedma2004}
Herrera-Viedma, E., Herrera, F., Chiclana, F., Luque, M.: {Some issues on
  consistency of fuzzy preference relations}.
\newblock Eur. J. Oper. Res. \textbf{154}(1), 98--109 (2004)

\bibitem{jannin2008}
Jannin, P., Korb, W.: Assessment of image-guided interventions.
\newblock In: T.~Peters, K.~Cleary (eds.) Image-Guided Interventions, pp.
  531--549. Springer US (2008)

\bibitem{Kranzfelder2013}
Kranzfelder, M., Schneider, A., Fiolka, A., Schwan, E., Gillen, S., Wilhelm,
  D., Schirren, R., Reiser, S., Jensen, B., Feu\ss~ner, H.: {Real-time
  instrument detection in minimally invasive surgery using radiofrequency
  identification technology}.
\newblock Journal of Surgical Research \textbf{185}(2), 704--710 (2013)

\bibitem{Lalys2014}
Lalys, F., Jannin, P.: {Surgical process modelling: a review.}
\newblock International journal of computer assisted radiology and surgery
  \textbf{9}(3), 495--511 (2014)

\bibitem{Merigo2010}
Merig\'{o}, J.M., Casanovas, M.: {Geometric operators in decision making with
  minimization of regret}.
\newblock Int. J. Comput. Syst. Sci. Eng. \textbf{1}(2), 111--118 (2007)

\bibitem{Neumuth2011}
Neumuth, T., Jannin, P., Schlomberg, J., Meixensberger, J., Wiedemann, P.,
  Burgert, O.: {Analysis of surgical intervention populations using generic
  surgical process models}.
\newblock Int. J. Comput. Assist. Radiol. Surg. \textbf{6}(1), 59--71 (2011)

\bibitem{Neumuth2012}
Neumuth, T., Liebmann, P., Wiedemann, P., Meixensberger, J.: {Surgical Workflow
  Management Schemata for Cataract Procedures. Process Model-based Design and
  Validation of Workflow Schemata.}
\newblock Methods of information in medicine \textbf{51}(4), 1--12 (2012)

\bibitem{Padoy2010}
Padoy, N., Blum, T., Ahmadi, S.A., Feussner, H., Berger, M.O., Navab, N.:
  Statistical modeling and recognition of surgical workflow.
\newblock Med. Image Anal. \textbf{16}(3), 632--641 (2012)

\bibitem{Padoy2011}
Padoy, N., Hager, G.D.: {Human-Machine Collaborative Surgery Using Learned
  Models}.
\newblock In: 2011 IEEE ICRA, pp. 5285--5292. IEEE (2011)

\bibitem{Perez2010}
Perez, I.J., Cabrerizo, F.J., Herrera-Viedma, E.: {A Mobile Decision Support
  System for Dynamic Group Decision-Making Problems}.
\newblock IEEE Trans. Syst., Man, Cybern. A, Syst., Humans \textbf{40}(6),
  1244--1256 (2010)

\bibitem{Saaty1990}
Saaty, T.L.: {How to make a decision: The analytic hierarchy process}.
\newblock Eur. J. Oper. Res. \textbf{48}(1), 9--26 (1990)

\bibitem{Saaty2008}
Saaty, T.L.: {Decision making with the analytic hierarchy process}.
\newblock Int. J. Services Sciences \textbf{1}(1), 83 (2008)

\end{thebibliography}

\end{document}